\documentclass[conference]{IEEEtran}
\IEEEoverridecommandlockouts
\usepackage{cite}
\usepackage{amsmath,amssymb,amsfonts}
\usepackage{algorithmic}
\usepackage{graphicx}
\usepackage{textcomp}
\usepackage{xcolor}
\usepackage{acronym}
\usepackage{booktabs}
\usepackage[caption=false]{subfig}

\renewcommand{\thetable}{\arabic{table}}

\newcommand{\eq}[1]{Eq.~\eqref{#1}}

\newcommand{\fig}[1]{Fig.~\ref{#1}}
\newcommand{\tab}[1]{Tab.~\ref{#1}}
\newcommand{\secref}[1]{Section~\ref{#1}}
\newcommand{\sys}[1]{System~(\ref{#1})}

\newcommand{\trg}{{\mathrm{trg}}}

\newcommand{\vv}[1]{\mathbf{#1}}

\graphicspath{{./figures/}}
\usepackage[colorlinks,linkcolor=black,citecolor=black,urlcolor=black]{hyperref}

\acrodef{prop}[\textit{MIMORPH}]{MIMO Radio Platform for Heterogeneous wireless systems}

\acrodef{3gpp}[3GPP]{3rd Generation Partnership Project}
\acrodef{abft}[A-BFT]{Association Beamforming Training}
\acrodef{ack}[ACK]{Acknowledgment}
\acrodef{adc}[ADC]{Analog-to-Digital Converter}
\acrodef{aoa}[AoA]{Angle of Arrival}
\acrodef{aod}[AoD]{Angle of Departure}
\acrodef{ap}[AP]{Access Point}
\acrodef{amc}[AMC]{Advanced Mezzanine Card}
\acrodef{awv}[AWV]{Antenna Wave Vector}
\acrodef{axi}[AXI]{Advanced eXtensible Interface}
\acrodef{ber}[BER]{Bit Error Rate}
\acrodef{bft}[BFT]{Beamforming Training}
\acrodef{bp}[BP]{Beam Pattern}
\acrodef{brp}[BRP]{Beam Refinement Phase}
\acrodef{cs}[CS]{Compressed Sensing}
\acrodef{cdf}[CDF]{Cumulative Distribution Function}
\acrodef{cef}[CEF]{Channel Estimation Field}
\acrodef{cfo}[CFO]{Carrier Frequency Offset}
\acrodef{sfo}[SFO]{Sampling Frequency Offset}
\acrodef{cir}[CIR]{Channel Impulse Response}
\acrodef{cfr}[CFR]{Channel Frequency Response}
\acrodef{csi}[CSI]{Channel State Information}
\acrodef{cs}[CS]{Compressed Sensing}
\acrodef{cv}[CV]{Constant Velocity}
\acrodef{cnn}[CNN]{Convolutional Neural Network}
\acrodef{cots}[COTS]{Commercial-Off-The-Shelf}
\acrodef{dft}[DFT]{Discrete Fourier Transform}
\acrodef{dl}[DL]{Deep Learning}
\acrodef{dma}[DMA]{Direct Memory Access}
\acrodef{dmg}[DMG]{Directional Multi Gigabit}
\acrodef{dti}[DTI]{Data Transfer Interval}
\acrodef{edmg}[EDMG]{Enhanced Directional Multi Gigabit}
\acrodef{ekf}[EKF]{Extended Kalman Filter}
\acrodef{kf}[KF]{Kalman Filter}
\acrodef{elu}[ELU]{Exponential-Linear Unit}
\acrodef{fmcw}[FMCW]{Frequency-Modulated Continuous-Wave}
\acrodef{fov}[FOV]{Field-of-View}
\acrodef{ft}[FT]{Fourier Transform}
\acrodef{fr2}[FR2]{Frequency Range 2}
\acrodef{gpio}[GPIO]{General Purpose Input/Output}
\acrodef{gsps}[GSPS]{Giga-Samples per Second}
\acrodef{har}[HAR]{Human Activity Recognition}
\acrodef{ht}[HT]{High Throughput}
\acrodef{idft}[IDFT]{Inverse Discrete Fourier Transform}
\acrodef{if}[IF]{Intermediate Frequency}
\acrodef{ifs}[IFS]{Inter-Frame Spacing}
\acrodef{iht}[IHT]{Iterative Hard Thresholding}
\acrodef{ista}[ISTA]{Iterative Shrinkage-Thresholding Algorithm}
\acrodef{isac}[ISAC]{Integrated Sensing And Communication}
\acrodef{jcs}[JCS]{Joint Communication and Sensing}
\acrodef{jpdaf}[JPDAF]{Joint Probabilistic Data Association Filter}
\acrodef{ls}[LS]{Least Squares}
\acrodef{lsd}[LSD]{Log-Spectral Distance}
\acrodef{los}[LoS]{Line-of-Sight}
\acrodef{lbm}[LBM]{Loop-Back Memory}
\acrodef{mae}[MAE]{Mean Absolute Error}
\acrodef{mcs}[MCS]{Modulation and Coding Scheme}
\acrodef{md}[$\mu$D]{micro-Doppler}
\acrodef{mimo}[MIMO]{Multiple Input Multiple Output}
\acrodef{mse}[MSE]{Mean Squared Error}
\acrodef{mmwave}[mmWave]{Millimeter-Wave}
\acrodef{msps}[MSPS]{Mega-Samples per Second}
\acrodef{mu}[MU]{Multiple User}
\acrodef{MUSIC}[MUSIC]{MUlti SIgnal Classification}
\acrodef{nac}[NAC]{Normalized Auto Correlation}
\acrodef{nco}[NCO]{Numerical Controlled Oscillator}
\acrodef{nlos}[NLoS]{Non-Line-of-Sight}
\acrodef{nn}[NN]{Neural Network}
\acrodef{nls}[NLS]{Nonlinear Least-Squares}
\acrodef{ofdm}[OFDM]{Orthogonal Frequency Division Multiplexing}
\acrodef{omp}[OMP]{Orthogonal Matching Pursuit}
\acrodef{per}[PER]{Packet Error Rate}
\acrodef{phy}[PHY]{Physical Layer}
\acrodef{pl}[PL]{Programmable Logic}
\acrodef{pov}[POV]{Point-of-View}
\acrodef{ps}[PS]{Processing System}
\acrodef{po}[PO]{Phase Offset}
\acrodef{pri}[PRI]{Pulse Repetition Interval}
\acrodef{psnr}[PSNR]{Peak Signal-to-Noise Ratio}
\acrodef{ransac}[RANSAC]{Random Sample Consensus}
\acrodef{rf}[RF]{Radio Frequency}
\acrodef{rfsoc}[RFSoC]{Radio Frequency System on a Chip}
\acrodef{rcs}[RCS]{Radar Cross-Section}
\acrodef{rss}[RSS]{Received Signal Strength}
\acrodef{rom}[ROM]{Read Only Memories}
\acrodef{rx}[RX]{receiver}
\acrodef{sc}[SC]{Single Carrier}
\acrodef{sdr}[SDR]{Software Defined Radio}
\acrodef{siso}[SISO]{Single Input Single Output}
\acrodef{sls}[SLS]{Sector Level Sweep}
\acrodef{snr}[SNR]{Signal-to-Noise Ratio}
\acrodef{ssnr}[SSNR]{Sensing Signal-to-Noise Ratio}
\acrodef{soc}[SoC]{System on a Chip}
\acrodef{spb}[SPB]{Signal Processing Blocks}
\acrodef{srrc}[SRRC]{Square-Root-Raised-Cosine}
\acrodef{ssr}[SSR]{Super Sample Rate}
\acrodef{sta}[STA]{Station}
\acrodef{std}[STD]{Standard Deviation}
\acrodef{stf}[STF]{Short Training Field}
\acrodef{stft}[STFT]{Short Time Fourier Transform}
\acrodef{su}[SU]{Single User}
\acrodef{tf}[TF]{Time-Frequency}
\acrodef{to}[TO]{Timing Offset}
\acrodef{toa}[ToA]{Time of Arrival}
\acrodef{tx}[TX]{Transmitter}
\acrodef{ula}[ULA]{Uniform Linear Array}
\acrodef{usrp}[USRP]{Universal Software Radio Peripheral}
\acrodef{vht}[VHT]{Very High Throughput}
\acrodef{wlan}[WLAN]{Wireless Local Area Network}
\acrodef{emd}[EMD]{Earth Mover's Distance}
\acrodef{6g}[6G]{6th Generation}
\acrodef{lfm}[LFM]{Linear Frequency Modulation}
\acrodef{fbmc}[FBMC]{Filter-Bank Multicarrier}
\acrodef{gfdm}[GFDM]{Generalized Frequency-Division Multiplexing}
\acrodef{dft-s-ofdm}[DFT-s-OFDM]{Discrete Fourier Transform-spread-OFDM}
\acrodef{otfs}[OTFS]{Orthogonal Time Frequency Space}
\acrodef{papr}[PAPR]{Peak-to-Average Power Ratio}
\acrodef{iot}[IoT]{Internet of Things}
\acrodef{rrss}[RRSS]{Received Raw Signal Sample}
\acrodef{csq}[CSQ]{Cellular Signal Quality}
\acrodef{pcl}[PCL]{Passive Coherent Location}
\acrodef{pslr}[PSLR]{Peak-to-Sidelobe Ratio}
\acrodef{isl}[ISL]{Integrated Sidelobe Level}
\acrodef{rd}[RD]{Range-Doppler}
\acrodef{api}[API]{Application Programming Interface}
\acrodef{agc}[AGC]{Automatic Gain Control}
\acrodef{crlb}[CRLB]{Cramér–Rao Lower Bound}
\acrodef{frft}[FRFT]{Fractional Fourier Transform}
\acrodef{osca}[OSCA]{Ordered-Statistic and Cell-Averaging}
\acrodef{cfar}[CFAR]{Constant False Alarm Rate}
\acrodef{doa}[DoA]{Direction of Arrival}
\acrodef{uav}[UAV]{Unmanned Aerial Vehicle}
\acrodef{awgn}[AWGN]{Additive White Gaussian Noise}
\acrodef{bw}[BW]{Bandwidth}
\acrodef{scs}[SCS]{Subcarrier Spacing}
\acrodef{cp}[CP]{Cyclic Prefix}
\acrodefplural{cp}[CPs]{Cyclic Prefixes}
\acrodef{pll}[PLL]{Phase-Locked Loop}

\acrodef{5g}[5G NR]{5G New Radio}
\acrodef{5ga}[5G-A]{5G-Advanced}
\acrodef{rs}[RS]{Reference Signal}
\acrodef{fr1}[FR1]{Frequency Range 1}
\acrodef{fr2}[FR2]{Frequency Range 2}
\acrodef{re}[RE]{Resource Element}
\acrodef{rg}[RG]{Resource Grid}
\acrodef{prb}[PRB]{Physical Resource Block}
\acrodef{rb}[RB]{Resource Block}
\acrodef{cran}[C-RAN]{Cloud Radio Access Network}

\acrodef{dmrs}[DM-RS]{Demodulation Reference Signal}
\acrodef{ptrs}[PT-RS]{Phase Tracking Reference Signal}
\acrodef{prs}[PRS]{Positioning Reference Signal}
\acrodef{csirs}[CSI-RS]{Channel-State Information Reference Signal}
\acrodef{pss}[PSS]{Primary Synchronization Signal}
\acrodef{sss}[SSS]{Secondary Synchronization Signal}
\acrodef{ssim}[SSIM]{Structural Similarity Index Metric}
\acrodef{ssb}[SSB]{Synchronization Signal Block}
\acrodef{rimrs}[RIM-RS]{Remote Interference Management-Reference Signal}
\acrodef{sib1}[SIB1]{System Information Block Type 1}
\acrodef{ss}[SS]{Synchronization Signal}

\acrodef{pbch}[PBCH]{Physical Broadcast Channel}
\acrodef{pdcch}[PDCCH]{Physical Downlink Control Channel}
\acrodef{pdsch}[PDSCH]{Physical Downlink Shared Channel}
\acrodef{pucch}[PUCCH]{Physical Uplink Control Channel}
\acrodef{pusch}[PUSCH]{Physical Uplink Shared Channel}

\acrodef{ue}[UE]{User Equipment}
\acrodef{gnb}[gNB]{gNodeB}

\acrodef{n1}[N1]{Node 1}
\acrodef{n2}[N2]{Node 2}
\acrodef{dbscan}[DBSCAN]{Density-Based Spatial Clustering of Applications with Noise}

\acrodef{tg}[TG]{Target}
\acrodef{rx}[RX]{Receiver}
\acrodef{rxn}[RX$_n$]{Receivers}
\acrodef{lo}[LO]{Local Oscillator}
\acrodef{t}[T]{Time}
\acrodef{path-one}[LoS]{Line of Sight}
\acrodef{path-two}[tgt]{target}
\acrodef{ifft}[IFFT]{Inverse Fast Fourier Transform}
\acrodef{mae}[MAE]{Mean Absolute Error}

\acrodef{imu}[IMU]{Inertial Measurement Unit}
\acrodef{ins}[INS]{Inertial Navigation System}
\acrodef{mems}[MEMS]{Micro-Electro-Mechanical System}
\acrodef{gnss}[GNSS]{Global Navigation Satellite System}
\acrodef{cog}[COG]{Course-over-Ground}
\acrodef{arw}[ARW]{Angle Random Walk}
\acrodef{vrw}[VRW]{Velocity Random Walk}
\acrodef{9dof}[9-DoF]{9 degrees of freedom}

\def\BibTeX{{\rm B\kern-.05em{\sc i\kern-.025em b}\kern-.08em
    T\kern-.1667em\lower.7ex\hbox{E}\kern-.125emX}}

\IEEEaftertitletext{\vspace{-0.5\baselineskip}}

\begin{document}

\bstctlcite{biblio:BSTcontrol}

\title{Reliable Doppler Estimation in Asynchronous Moving ISAC Devices via IMU Integration}
\author{Zaman Bhalli$^{\dag}$, Jacopo Pegoraro$^{\dag}$, Gianmaria Ventura$^{\dag}$,  Paolo Casari$^{\ddag}$, Michele Rossi$^{\dag * }$
\thanks{$^{\dag}$ These authors are with the University of Padova, Dept. of Information Engineering, Italy. $^{*}$ This author is with the University of Padova, Dept. of Mathematics, Italy.
$^{\ddag}$ This author is with DISI, University of Trento, Italy.
The work presented in this paper is supported by the Multi-X project that has received funding from the Smart Networks and Services Joint Undertaking (SNS JU) under the European Union’s Horizon Europe research and innovation programme under Grant Agreement No 101192521. 
Corresponding author email: \texttt{zaman.bhalli@unipd.it}. 
}
}

\maketitle

\begin{abstract}
We present an \ac{isac} framework for the joint estimation of the Doppler frequency of a passive mobile target in a bistatic scenario with clock-asynchronous nodes and where the \ac{rx} is static but the \ac{tx} is mobile. In such a setup, previous geometric solutions are integrated with an \ac{imu} device at the \ac{tx}, coming up with a truly joint data fusion and estimation algorithm based on extended Kalman filtering. The approach jointly estimates the target Doppler frequency, along with the speed and direction of motion of the \ac{tx} and it is robust to the unavailability of static paths between the \ac{tx} and \ac{rx} pair, a condition that makes previous solutions ineffective. The developed extended Kalman filter strikes a balance between the accuracy of \ac{isac} and the \ac{imu} reliability. The proposed solution is validated via numerical simulations, obtaining a median Doppler error of 1.2\% with a smartphone-grade \ac{imu} under realistic operating conditions.
\end{abstract}


\section{Introduction}
\label{sec:introduction}

Considered one of the cornerstones of the \ac{6g} of wireless networks~\cite{imt-2030}, \ac{isac} embeds radio-based sensing  directly into communication systems. 
In this paper, we focus on the estimation of the Doppler shift of a passive target~\cite{canil2022milliTRACE-IR}. Specifically, we consider a bistatic configuration, where the independent clocks of the \ac{tx} and \ac{rx} give rise to \ac{to}, \ac{cfo}, and \ac{po}. In \ac{isac} systems, these act as disturbance terms and hinder the coherent processing of the channel estimates over time~\cite{wu2024sensing}. Moreover, we consider that the \ac{tx} is mobile. This introduces additional complexity, as the \ac{tx} motion induces an additional Doppler contribution on every propagation path, which combines in a non-trivial manner with that of the target.

Current studies mainly deal with asynchronous clocks by assuming \emph{static} devices~\cite{pegoraro2024jump, li2022csiratio}, or consider \emph{moving} devices in monostatic-like arrangements, which are not affected by the aforementioned timing, carrier, and phase offsets~\cite{zhang2022mobi2sense}. In the literature, the Doppler estimation in bistatic setups is dealt with by processing the phases of the multipath components of the \ac{cir}, either with a single \ac{rx} conditioned by external static scatterers~\cite{ventura2024bistatic, ventura2025asymov}, or with multiple \acp{rx} and no external reflectors~\cite{bhalli2026multistatic}. These works propose geometric solutions that are solvable when enough reference  paths are concurrently available, i.e., at least two static scatterers in~\cite{ventura2024bistatic} and~\cite{ventura2025asymov}, or at least four \acp{rx} in~\cite{bhalli2026multistatic}. In real environments, the scatterers appear and disappear as the scene evolves, leading to frequent {\it measurement dropouts}~\cite{sinopoli2004kalman}.

In the present work, we tackle this issue by augmenting geometric solutions with an inertial sensor. In fact, low-cost  \acp{imu} devices are available in virtually every \ac{ue}-class device and provide self-contained motion estimates at a reasonable rate. Their open-loop integration, however, accumulates the sensor biases and drifts unboundedly with time~\cite{titterton2004strapdown}. Tightly coupled radar-inertial filters are well established in robotic navigation~\cite{doer2020ekf}, but rely on monostatic radars. The closest \ac{imu} integration attempt within \ac{isac} is that of~\cite{ventura2025asymov}, where the \ac{imu} speed measurement is plugged into the \ac{isac} system {\it as a known quantity}, reducing the resolvability requirement to a single static scatterer. 

In this study, we take the \ac{isac} framework of~\cite{ventura2024bistatic, ventura2025asymov} as our starting point, by integrating an \ac{imu} at the mobile \ac{tx}, towards tracking the Doppler frequency of a moving passive target. We consider a single static \ac{rx} and a \ac{tx} (acting as a mobile \ac{ue}) that carries the \ac{imu}, while up to $N_{\rm s}$ static scatterers are present only a random fraction of the time. Rather than solving the resulting equations frame by frame, we treat them, together with the \ac{imu} measurements, as the measurement rows of a tightly-coupled \ac{ekf}: the drift-free \ac{isac} rows re-anchor the accumulated \ac{imu} drift whenever a scatterer returns, while the always-available \ac{imu} rows sustain the estimation when the static scatterers are not available. Different from~\cite{ventura2025asymov}, \ac{isac} and \ac{imu} {\it correct each other} instead of the \ac{imu} acting as a substitute when static paths become unavailable. Our main contributions are:
\begin{enumerate}
    \item We extend the framework of~\cite{ventura2024bistatic, ventura2025asymov} by integrating an \ac{imu}, jointly tracking the target Doppler frequency and the \ac{tx} device speed and orientation in bistatic \ac{isac} with asynchronous nodes through an \ac{ekf}.
    \item We consider the challenging but realistic case where the static scatterers are only intermittently visible. The developed \ac{ekf} accepts any number of scatterer paths, including none, and produces estimates where the existing solutions fail.
    \item We evaluate the proposed solution via numerical simulations and characterize the impact of \ac{aod} noise, node speeds, \ac{imu} grades, frame period, and scatterer visibility.
\end{enumerate}


\section{System Model}
\label{sec:system-model}


\subsection{Reference Scenario}
\label{sec:ref-scenario}

\begin{figure}
  \centering
  \includegraphics[width=\columnwidth]{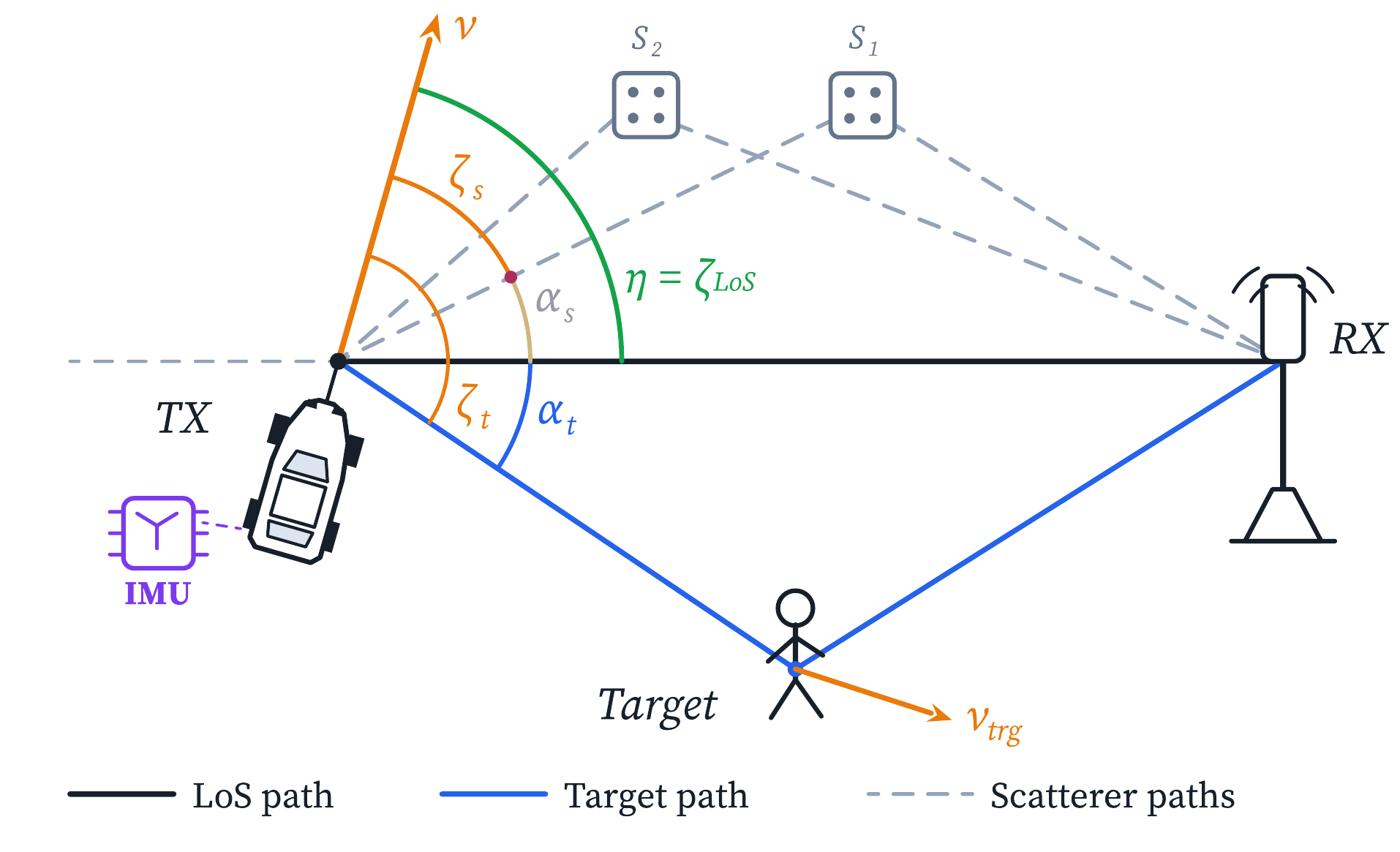}
      \caption{Geometric representation of the considered scenario showing \ac{los}, target and intermittent scatterer paths, and the \ac{imu} mounted on the \ac{tx}.}
  \label{geometry}
  \vspace{-0.3cm}
  \end{figure}

We consider a 2D asynchronous bistatic \ac{isac} scenario with a single mobile \ac{tx}, one static \ac{rx}, a moving target, and $N_{\rm s}$ static scatterers acting as environmental anchors (whose presence is not guaranteed), indexed by $s = 1, \dots, N_{\rm s}$.
A~low-cost \ac{imu} is mounted on the \ac{tx} as shown in~\fig{geometry}.
The \ac{tx} moves with velocity vector $\vv{v} \in \mathbb{R}^2$, having magnitude $v = \|\vv{v}\|$ and orientation $\eta$, the latter measured with respect to the \ac{tx}--\ac{rx} direction, while the target moves with an independent velocity vector $\vv{v}_{\trg}\in \mathbb{R}^2$ of magnitude $v_{\trg}$.

The \ac{rx} observes the following components in the multipath wireless channel: \emph{(i)}~the Line-of-Sight (LoS) path, \emph{(ii)}~a \emph{target} path scattered off the moving target, and \emph{(iii)}~up to $N_{\rm s}$
\emph{scatterer} paths, one per scatterer, emanating from the static anchors. 
When none of the scatterers is visible, the multipath environment reduces to two resolvable paths, the \ac{los} and the target, which we assume to be always available. 

Because the \ac{tx} and the \ac{rx} are driven by independent clocks, every path is impaired by \ac{cfo}, \ac{po}, and \ac{to}~\cite{wu2024sensing}. A key property exploited throughout this work is that these offset terms are \emph{common to all multipath components} at the \ac{rx} for a particular unit time.
Besides the Doppler frequency of the target, which is the sensing quantity of interest, the \ac{tx} motion is itself an unknown: its velocity and orientation induce a Doppler contribution on \emph{every} path, which is entangled with that of the target and must be resolved.

\subsection{Channel Model}
\label{sec:channel_model}

Practical implementations of \ac{isac} systems require the discretization of continuous-time signals. 
In our system, the \ac{rx} estimates the \ac{cfr} from each \ac{5g} \ac{dmrs} pilot received and converts it into a \ac{cir} estimate through \ac{idft}. 
The estimation is repeated with period $T$ and indexed by $k=0, \dots, K-1$. 
Under the common assumption that the channel's dynamic characteristics change negligibly throughout the short interval $T$, we discretize the continuous-time \ac{cir} in discrete time $k$ and delay bin $\ell$ as~\cite{mimo_radar}
\begin{equation}
    \begin{split}
    h[k, \ell] =  e^{j\psi^{\mathrm{o}}[k]} \hspace{-0.2cm}\sum_{m \in \mathcal{M}[k]}\hspace{-0.2cm} A_{m}[k] e^{j\vartheta_{m}[k]}  \delta\bigl[\ell \hspace{-0.05cm}- \hspace{-0.05cm}\tau_{m}\hspace{-0.05cm} - \hspace{-0.05cm} \tau^{\mathrm{o}}[k]\bigr] + w[k, \ell],
    \end{split}
    \label{eq:cir-discrete}
\end{equation}
where $A_{m}[k]$ is the complex path amplitude, $\tau_{m}$ the propagation delay, $\psi^{\mathrm{o}}[k]$ the \ac{po}, $\tau^{\mathrm{o}}[k]$ the \ac{to}, $w[k,l]$ an \ac{awgn} term.
Index $m$ spans the set $\mathcal{M}[k] = \{\mathrm{LoS}, \trg\} \cup \mathcal{S}[k]$ of the visible paths at frame $k$, i.e., the \ac{los} and the target paths ($\mathrm{LoS}$ and $\trg$), plus the scatterer paths collected in $\mathcal{S}[k] \subseteq \{1, \dots, N_{\rm s}\}$, with $N_{\rm s}[k] = |\mathcal{S}[k]| \leq N_{\rm s}$. 
Note that only the \ac{los} and target paths need to be identified and distinguished from the static paths.
The intermittent static components in $\mathcal{S}[k]$ can be used interchangeably by the \ac{rx}.

Hereafter, $m$ denotes a generic path, $s \in \mathcal{S}[k]$ a scatterer, and $\bar{\mathcal{M}}[k] = \mathcal{M}[k] \smallsetminus \{\mathrm{LoS}\}$.
The phase term is
\begin{equation}
    \vartheta_{m}[k] = 2\pi kT \left(f_{m}^{\mathrm{tx}}[k] + f_{\rm d}[k] + f^{\mathrm{o}}[k]\right),
    \label{eq:phase-components-discrete}
\end{equation}
 where $f_{m}^{\mathrm{tx}}[k]$ and $f_{\rm d}[k]$ denote the \ac{tx}'s Doppler frequency on path $m$ and the target's Doppler frequency, respectively, the latter being non-zero on the target path only, and $f^{\mathrm{o}}[k]$ is the \ac{cfo}. 
 As anticipated in \secref{sec:ref-scenario}, $f^{\mathrm{o}}[k]$ does not depend on the path index $m$ and is common to all paths.
 
\subsection{Doppler Geometry}
\label{sec:doppler-geometry}

We now relate the Doppler terms in \eq{eq:phase-components-discrete} to the propagation geometry, taking \fig{geometry} as a reference. 
Since the \ac{tx} is the common origin of all propagation paths, its motion induces a different Doppler shift on each of them. We denote by $\zeta_m[k]$ the angle between the \ac{tx} velocity vector $\vv{v}$ and the segment connecting the \ac{tx} to the scatterer of the $m$-th path, $m \in \mathcal{M}[k]$, at time $kT$.

The \ac{tx}-induced Doppler shift on path $m$ is~\cite{mimo_radar} 
\begin{equation}
    f_{m}^{\mathrm{tx}}[k] = \frac{{v}[k]}{\lambda}\,\cos(\zeta_{m}[k]),
    \label{eq:tx-doppler}
\end{equation}
where $\lambda = c/f_{\rm c}$ is the carrier wavelength, $c$ is the speed of light and $f_{\rm c}$ the carrier frequency. For the \ac{los} path the segment connects the \ac{tx} to the \ac{rx}, which is the reference direction of the \ac{tx} orientation (see \secref{sec:ref-scenario}), hence $\zeta_{\mathrm{LoS}}[k] = \eta[k]$. The target-induced term $f_{\rm d}[k]$, instead, is non-zero on the target path only.
Following~\cite{ventura2024bistatic}, we do not expand $f_{\rm d}$ into its geometric components and treat it as a single quantity. In the remainder of this work, $f_{\rm d}$, ${v}$, and $\eta$ are the unknowns to be jointly estimated.

\subsection{Measured Phase Model}
\label{sec:phase-model}

The estimation methodology of this work (\secref{sec:methodology}) mainly exploits the phases corresponding to the \ac{los}, target, and scatterer paths (if available) in the \ac{cir}. From \eqref{eq:cir-discrete}, the phase of the generic $m$-th path at time $k$ is
\begin{equation}
  \label{eq:phase-model}
  \phi_{m}[k] = \Psi^{\mathrm{o}}[k] + \angle A_{m} + \chi_{m}[k] + \epsilon_{m}[k],
\end{equation}
where $\Psi^{\mathrm{o}}[k] = \psi^{\mathrm{o}}[k] + 2\pi kT f^{\mathrm{o}}[k]$ summarizes the combined effect of \ac{po} and \ac{cfo}, which is common to all paths at time $k$, $\angle \cdot$ is the phase operator, $\chi_{m}[k] = 2\pi k T \left(f_{m}^{\mathrm{tx}} + f_{\mathrm{d}}\right)$ captures the contribution of the Doppler components as mentioned in \secref{sec:doppler-geometry}, and $\epsilon_{m}[k]$ is a residual noise term. $\angle A_{m}$ indicates the phase component due to the length of the path. Considering a short-enough time window, it is reasonable to assume that $\angle A_{m}$ remains constant across different time instants, and we therefore drop its time index. As previously discussed, the scatterer paths obey \eq{eq:phase-model} as well but, different from the \ac{los} and the target paths, which are always present, they are intermittently available, i.e., they appear only at those times $k$ in which the corresponding scatterer belongs to $\mathcal{S}[k]$.

\subsection{IMU Modeling}
\label{sec:imu-model}

We assume the \ac{tx} to have a \ac{9dof} \ac{imu} (gyroscope, accelerometer, and magnetometer) rigidly mounted on the device~\cite{titterton2004strapdown}. 
Under a typical (e.g., pedestrian, vehicular, etc.) 2D motion model where the longitudinal axis aligns with the velocity vector, the \ac{imu} measures the yaw rate $\omega(t) = \partial \eta(t)/\partial t$ and longitudinal acceleration $a(t) = \partial v(t) /\partial t$, with $v$ and $\eta$ representing the \ac{tx} velocity magnitude and direction of \secref{sec:ref-scenario}.

The \ac{imu} acquires samples with period $T_{\rm imu}$ on the discrete time grid $i\,T_{\rm imu}$, $i \in \mathbb{N}$. 
Real measurements are corrupted by thermo-mechanical noise and slow-varying biases~\cite{titterton2004strapdown}. 
The discrete-time sample $i$ is thus modeled as
\begin{equation}
    \tilde{\omega}_i = \omega_i + b_{\omega,i} + n_{\omega,i}, \quad
    \tilde{a}_i = a_i + b_{a,i} + n_{a,i},
    \label{eq:imu-noise}
\end{equation}
where $n_{\omega,i} \sim \mathcal{N}(0,\sigma_{\omega}^2)$ and $n_{a,i} \sim \mathcal{N}(0,\sigma_{a}^2)$ are i.i.d.\ white Gaussian noise processes. The bias terms follow a random walk $b_{\omega,i} = b_{\omega,i-1} + u_{\omega,i}$ with $u_{\omega,i}\sim\mathcal{N}(0,\sigma_{b,\omega}^2)$ (and analogously for $b_{a,i}$ with $\sigma_{b,a}^2$).

The quadruple $(\sigma_\omega, \sigma_a, \sigma_{b,\omega}, \sigma_{b,a})$ fully determines the quality of the \ac{imu}. 
We tie these four terms to the gyroscope noise by setting $\sigma_a = 10\,\sigma_\omega$ and $\sigma_{b,x} = \sigma_{x}/10$ with $x\in\{\omega, a\}$, so that a single value of $\sigma_\omega$ identifies the quality of the \ac{imu}.
Open-loop integration from an initial pair of values $({v}_0, {\eta}_0)$ yields estimates ${\eta}^{\rm imu}_i$ and ${v}^{\rm imu}_i$ on the \ac{imu} grid, which we interpolate onto the frame instants to obtain $\eta^{\rm imu}[k]$ and $v^{\rm imu}[k]$.
The initialization values could be obtained by using the on-board magnetometer and \ac{gnss} over a short boot interval of $T_{\rm init}$ preceding the estimation. 
Throughout this interval, the on-board magnetometer and the \ac{gnss} speed-over-ground deliver, at each of their samples, noisy readings of the true orientation and speed, modeled as $\tilde{\eta}_0 = \eta_0 + n_{\rm mag}$ and $\tilde{v}_0 = v_0 + n_{\rm gnss}$,
with $n_{\rm mag}\sim\mathcal{N}(0,\sigma_{\rm mag}^2)$ and $n_{\rm gnss}\sim\mathcal{N}(0,\sigma_{\rm gnss}^2)$. Averaging the readings collected over $T_{\rm init}$ attenuates these errors, yielding an initial pair of estimates $(\hat{v}_0,\hat{\eta}_0)$. 


\subsection{Noise Characterization}
\label{sec:noise-model}

The two modalities, i.e., \ac{isac} and \ac{imu}, carry errors of a completely different nature, that
the fusion filter that will be presented in \secref{sec:fusion-ekf} has to weigh against one another. 

\subsubsection{ISAC Phase Noise}
The residual term $\epsilon_{m}[k]$ of \eq{eq:phase-model} stems from the noise of the \ac{cir} estimate, and it is modeled as $\epsilon_{m}[k] \sim \mathcal{N}(0, \sigma_{\phi,m}^2)$. The model is derived in~\cite{ventura2025asymov} for this same phase processing, with variance 
\begin{equation}
  \sigma_{\phi,m}^{2}[k] = \frac{\sigma_{h}^{2}}{2\,|A_{m}[k]|^{2}}
   = \frac{\sigma_{w}^{2}}{2\,G\,|A_{m}[k]|^{2}},
  \label{eq:path-phase-var}
\end{equation}
where $\sigma_{h}^{2}$ is the variance of the \ac{cir} estimation error,
$\sigma_{w}^{2}$ the noise power on the received signal, and $G$ the \ac{snr} gain of the
channel estimation, so that $\sigma_{h}^{2} = \sigma_{w}^{2}/G$. Note how a stronger path
carries a cleaner phase: the extracted phases are not equally reliable, and
\eq{eq:path-phase-var} quantifies by how much. Since the scale of the estimated \ac{cir}
is arbitrary, only amplitude ratios are identifiable, hence we anchor
\eq{eq:path-phase-var} to the \ac{los} and obtain every other path from the peaks,
\begin{equation}
  \sigma_{\phi,m}[k] = \sigma_{\phi,\rm LoS}\,
  \frac{|A_{\rm LoS}[k]|}{|A_{m}[k]|}, \qquad
  \sigma_{\phi,\rm LoS}^{2} = \frac{1}{2\,G\,\mathrm{SNR}},
  \label{eq:sigma-anchor}
\end{equation}
with $\mathrm{SNR} = |A_{\rm LoS}|^{2}/\sigma_{w}^{2}$ being the communication
\ac{snr} measured on the \ac{los}. 

\subsubsection{IMU Noise}
The \ac{imu} measurements $\eta^{\rm imu}[k]$ and $v^{\rm imu}[k]$ are open-loop integrals of the \ac{imu} equations \eq{eq:imu-noise}, initiated with the pair $(\hat{v}_0,\hat{\eta}_0)$. 
Their error is the error of this initial estimate plus an accumulated error term that increases with time. 
This process is described by a biased random walk on $\eta$ and~$v$, with variances 
\begin{equation}
  \begin{split}
  \sigma_{\eta}^{2}(t) &= \sigma_{\eta,0}^{2}
     + \sigma_{\omega}^{2}\,T_{\rm imu}\,t
     + \frac{\sigma_{b,\omega}^{2}\,t^{3}}{3\,T_{\rm imu}}, \\
  \sigma_{v}^{2}(t) &= \sigma_{v,0}^{2}
     + \sigma_{a}^{2}\,T_{\rm imu}\,t
     + \frac{\sigma_{b,a}^{2}\,t^{3}}{3\,T_{\rm imu}},
  \end{split}
  \label{eq:imu-cov}
\end{equation}
where $t$ is the time since the last fix, whose accuracies are $\sigma_{\eta,0}=\sigma_{\rm mag}/\sqrt{N_{\rm mag}}$ and $\sigma_{v,0}=\sigma_{\rm gnss}/\sqrt{N_{\rm gnss}}$, $N_{x}$ represents the number of samples each
sensor averages over $T_{\rm init}$. Since the gyroscope and the accelerometer are separate sensors with independent noise, and the planar open-loop trajectory integration keeps $\eta^{\rm imu}$ and $v^{\rm imu}$ on separate integration chains, the two errors of \eq{eq:imu-cov} are mutually uncorrelated.

\section{Methodology}
\label{sec:methodology}

\begin{figure}
  \centering
  \includegraphics[width=\columnwidth]{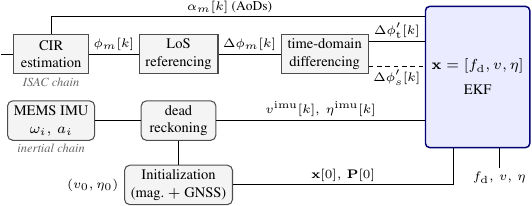}
  \caption{Block diagram of the proposed methodology. The \ac{isac} and \ac{imu} chain produces the differential phases $\Delta\phi'_{m}[k]$ and $v^{\rm imu}[k]$, $\eta^{\rm imu}[k]$, respectively. Both are fed to the \ac{ekf} for tracking $\vv{x} = [f_{\rm d}, v, \eta]^{\mathsf{T}}$}
  \label{block_diagram}
  \vspace{-0.3cm}
  \end{figure}

The proposed methodology comprises two stages, summarized in \fig{block_diagram}. The
first produces the \ac{isac} and \ac{imu} measurements: the \ac{isac} chain reduces the estimated \ac{cir} to phase measurements while the \ac{imu} chain produces the \ac{imu} samples via the open-loop integration of \secref{sec:noise-model}. The second stage, which is the core contribution of this work, fuses \ac{isac} and \ac{imu} measurements through an \ac{ekf}.

\subsection{ISAC Measurements}
\label{sec:isac-meas}
Let $\phi_{m}[k]$ be the phase of the $m$-th \ac{cir} peak at frame $k$. 
We reference every path to the \ac{los} and difference two consecutive
frames, i.e., $\Delta\phi_{m}[k] = \phi_{m}[k] - \phi_{\mathrm{LoS}}[k]$ and
$\Delta\phi'_{m}[k] = \Delta\phi_{m}[k] - \Delta\phi_{m}[k-1]$. 
The former cancels the \ac{po} and the \ac{cfo}, common to all paths of a frame, at the cost of the \ac{los} equation, the latter the path-length phase, assumed constant over a short window~\cite{pegoraro2024jump}. 
Of the $N_{\rm s}[k] + 2$ extracted phases, $N_{\rm s}[k] + 1$ remain useful.

The per-path angles $\zeta_{m}$ of \eq{eq:tx-doppler} are conveniently re-expressed using $\eta$ through the measured \acp{aod}. Denoting by $\alpha_{m}[k]$ the \ac{aod} of path $m$ with respect to the \ac{tx}--\ac{rx} direction, since $\alpha_{\mathrm{LoS}}[k] = 0$ by construction, geometric considerations lead to $\cos\zeta_{m}[k] = \cos(\alpha_{m}[k] - \eta[k])$. This yields the working equation of this paper,

\begin{equation}
  \begin{split}
  \Delta\phi'_{\trg}[k] &= 2\pi T\bigl[f_{\rm d}[k]
  + g_{\trg}\!\left(v[k], \eta[k]\right)\bigr], \\
  \Delta\phi'_{s}[k] &= 2\pi T\, g_{s}\!\left(v[k], \eta[k]\right),
  \quad s \in \mathcal{S}[k],
  \end{split}
  \label{eq:phase-row}
\end{equation}
where the target Doppler appears in the target row only, and
\begin{equation}
  g_{m}(v, \eta) = \frac{v}{\lambda}\left[\cos(\alpha_{m} - \eta)
  - \cos\eta\right], \quad m \in \bar{\mathcal{M}}[k]
  \label{eq:g-function}
\end{equation}

\sys{eq:phase-row} thus counts $N_{\rm s}[k] + 1$ equations in $3$ unknowns $f_d$, $v$, $\eta$ and is solvable within each frame $k$ only if \eqref{eq:phase-row} has at least $3$ equations, i.e., if $N_{\rm s}[k] \geq 2$, which is the operational condition of~\cite{ventura2024bistatic, ventura2025asymov}. This is precisely the weaknesses that the \ac{imu} covers, motivating the fusion filter presented next.

\subsection{ISAC--IMU Fusion via Extended Kalman Filter}
\label{sec:fusion-ekf}

The second stage tracks the state $\vv{x} = [f_{\rm d},\, v,\, \eta]^{\mathsf{T}}$ over time through an \ac{ekf} fusing \eq{eq:phase-row} with the \ac{imu} measurements ${v}^{\rm imu}[k]$ and ${\eta}^{\rm imu}[k]$. The filter is initialized with $(\hat{v}_0, \hat{\eta}_0)$ of \secref{sec:imu-model} and $f_{{\rm d},0} = 0$, with the initial covariance $\vv{P}$ matching the accuracy of that fix through $\sigma_{v,0}$ and $\sigma_{\eta,0}$ of \eq{eq:imu-cov}, and kept deliberately large on $f_{\rm d}$ to cover the whole Doppler range compatible with the operating speeds. Since the \ac{imu} already carries the \ac{tx} dynamics, modeling them in the transition as well would duplicate the same source of information. We therefore adopt a random-walk state evolution, so that the \emph{Prediction} step leaves the state unchanged and only inflates the state covariance as $\vv{P}[k|k\!-\!1] = \vv{P}[k\!-\!1] + \vv{Q}$, the process noise $\vv{Q}$ scaling with $T$ so that the filter adapts to the sampling of the \ac{isac} and \ac{imu} streams.

The \emph{Update} step collects the $3 + N_{\rm s}[k]$ scalar measurements available at frame $k$, i.e., the two \ac{imu}, the $N_{\rm s}[k] + 1$ \ac{isac} measurements. Hence, the measurement model of the filter is
\begin{equation}
  \renewcommand*{\arraystretch}{1.15}
  \underbrace{
  \begin{bmatrix}
    \eta^{\rm imu} \\
    v^{\rm imu} \\
    \Delta\phi'_{1} \\
    \vdots \\
    \Delta\phi'_{N_{\rm s}[k]} \\
    \Delta\phi'_{\trg}
  \end{bmatrix}}_{\vv{z}[k]}
  =
  \underbrace{
  \begin{bmatrix}
    \eta \\
    v \\
    2\pi T\, g_{1}(v, \eta) \\
    \vdots \\
    2\pi T\, g_{N_{\rm s}[k]}(v, \eta) \\
    2\pi T\bigl[ f_{\rm d} + g_{\trg}(v, \eta)\bigr]
  \end{bmatrix}}_{h(\vv{x}[k])}
  +
  \underbrace{
  \begin{bmatrix}
    n_{\eta} \\
    n_{v} \\
    n_{\phi,1} \\
    \vdots \\
    n_{\phi,N_{\rm s}[k]} \\
    n_{\phi,\trg}
  \end{bmatrix}}_{\vv{n}[k]},
  \label{eq:meas-model}
\end{equation}
where $\eta^{\rm imu}, v^{\rm imu}$ are the \ac{imu} readings while the true $\eta, v$ are the state entries. $N_{\rm s}[k]$ intermediate rows refer to the scatterers $s \in \mathcal{S}[k]$ visible at frame $k$, and $\vv{n}[k] \sim \mathcal{N}(\vv{0}, \vv{R}[k])$.

The two groups of rows in \eq{eq:meas-model} are produced by the two physical chains of \secref{sec:noise-model}, which {\it share no source of randomness}, hence $
  \vv{R}[k] = \mathrm{blkdiag}\bigl(\sigma_{\eta}^{2}(t_{k}),\;\sigma_{v}^{2}(t_{k}),\;
  \vv{C}[k]\bigr),
$ 
where $t_{k}$ is the time elapsed since the boot fix and $\vv{C}[k]$ has size $(N_{\rm s}[k]+1)\times(N_{\rm s}[k]+1)$, so that $\vv{R}[k]$ adapts to the scatterers that are available (visible) at frame $k$. The two \ac{imu} entries are provided by \eq{eq:imu-cov} evaluated at $t_{k}$.

The \ac{isac} rows, instead, are correlated. Propagating \eq{eq:path-phase-var} through the phase processing of \secref{sec:isac-meas} as in~\cite[\S IV-C]{ventura2025asymov}, the entries of the \ac{isac} block read
\begin{equation}
  C_{mm}[k] = \frac{\sigma_{w}^{2}\,\kappa_{m}[k]}{G}, \qquad
  C_{ml}[k] = \frac{\sigma_{w}^{2}}{G\,|A_{\rm LoS}[k]|^{2}},
  \label{eq:isac-cov}
\end{equation}
with $m \neq l$ and $\kappa_{m}[k] = |A_{m}[k]|^{-2} + |A_{\rm LoS}[k]|^{-2}$, $\vv{C}[k]$ becomes $2\,\mathrm{diag}(\sigma_{\phi,m}^{2}[k]) + 2\,\sigma_{\phi,\rm LoS}^{2}[k]\,\vv{1}\vv{1}^{\mathsf{T}}$, where $\vv{1}$ is all-ones vector of length $N_{\rm s}[k]+1$.
We calibrate $\sigma_{\phi,\rm LoS}$, the only free parameter left by \eq{eq:sigma-anchor}.
The \ac{imu} block grows monotonically as the trajectory integration accumulates error, whereas the \ac{isac} one merely follows the peak magnitudes, which change as slowly as the geometry does.

The measurement model is nonlinear in $(v, \eta)$ through $g_{m}(v,\eta)$, which prevents a direct application of the linear Kalman filter: the \ac{ekf} handles it by linearizing $h(\vv{x})$ around the current state estimate through the Jacobian $\vv{H} = \partial h(\vv{x}) / \partial \vv{x}$. Differentiating \eq{eq:meas-model}, and letting $c_{m} = \cos(\alpha_{m} - \eta) - \cos\eta$ and $d_{m} = \sin(\alpha_{m} - \eta) + \sin\eta$ for notational convenience, we obtain

\begin{equation}
  \renewcommand*{\arraystretch}{1.3}
  \vv{H} =
  \begin{bmatrix}
    0 & 0 & 1 \\
    0 & 1 & 0 \\
    0 & \frac{2\pi T\, c_{1}}{\lambda} & \frac{2\pi T\, v\, d_{1}}{\lambda} \\
    \vdots & \vdots & \vdots \\
    0 & \frac{2\pi T\, c_{N_s[k]}}{\lambda} & \frac{2\pi T\, v\, d_{N_s[k]}}{\lambda} \\
    2\pi T\, & \frac{2\pi T\, c_{\trg}}{\lambda} & \frac{2\pi T\, v\, d_{\trg}}{\lambda}
  \end{bmatrix},
  \label{eq:jacobian}
\end{equation}
whose first two rows state that the \ac{imu} observes $\eta$ and $v$ directly, while the last one is the only row observing $f_{\rm d}$. All rows are evaluated at the current state estimate.

The filter applies all the rows of a frame in a single batch update. Let
$\vv{\nu}[k] = \vv{z}[k] - h(\hat{\vv{x}}[k])$ be the innovation vector, wrapped to
$(-\pi,\pi]$ on the $\eta$ and phase rows. 
Then, the \ac{ekf} update computes
\begin{equation}
  \begin{split}
  \vv{S}[k] &= \vv{H}\vv{P}\vv{H}^{\mathsf{T}} + \vv{R}[k], \qquad
  \vv{K}[k] = \vv{P}\vv{H}^{\mathsf{T}}\vv{S}[k]^{-1}, \\
  \vv{x} &\leftarrow \vv{x} + \vv{K}[k]\,\vv{\nu}[k], \qquad\;\;
  \vv{P} \leftarrow \bigl(\vv{I}_{3} - \vv{K}[k]\vv{H}\bigr)\vv{P}.
  \end{split}
  \label{eq:batch-update}
\end{equation}
Processing the frame as a
whole is what the correlation of \eq{eq:isac-cov} requires. Applying the rows one at a
time as scalar corrections would be legitimate only if they were mutually independent,
which the shared \ac{los} noise prevents. This architecture still accommodates dynamic
measurement dimensions and arbitrary observation dropouts without structural
re-initialization~\cite{sinopoli2004kalman}, since a visibility outage simply removes
rows from \eq{eq:meas-model} and shrinks $\vv{C}[k]$ accordingly, while the ever-present
\ac{imu} rows keep the state observable throughout. 

\section{Numerical Results and Analysis}
\label{sec:results}


\subsection{Simulation Setup}
\label{sec:sim-setup}

\begin{table}[t]
    \centering
        \caption{Simulation parameters.}
    \setlength{\tabcolsep}{1.5pt}
    \begin{tabular}{l c l c}
        \toprule
        \multicolumn{2}{c}{\textit{Default System \& geometry}} &
        \multicolumn{2}{c}{\textit{Default Noise, \acs{imu} \& \acs{ekf}}} \\
        \cmidrule(lr){1-2}\cmidrule(lr){3-4}
        Parameter & Value & Parameter & Value \\
        \midrule
        Carrier $f_{\rm c}$      & $28$~GHz                 & \acs{snr}                                    & $20$~dB \\
        \acs{scs} / \acsp{rb}    & $120$~kHz / $52$         & \acs{aod} noise $\sigma_{\alpha}$            & $5^{\circ}$ \\
        Frame period $T$         & $125~\mu$s               & \acs{los} phase std $\sigma_{\phi,\rm LoS}$ & $0.5^{\circ}$ \\
        \acs{tx} power           & $200$~mW                 & \acs{imu} period $T_{\rm imu}$               & $1$~ms \\
        \acs{tx} speed / $\eta$  & $5$\,m/s / $315^{\circ}$ & Gyro noise $\sigma_{\omega}$                 & $10^{-2}$~rad/s \\
        Target speed / angle     & $5$\,m/s / $45^{\circ}$  & Accel.\ noise $\sigma_{a}=10\sigma_{\omega}$ & $0.1$~m/s$^2$ \\
        Scatterers $N_{\rm s}$ / vis.  & $2$ / $35\%$             & Bias walk $\sigma_{b,\bullet}$               & $\sigma_{\bullet}/10$ \\
        Burst length             & $2$--$20$ frames         & 
        $\sigma_{\rm mag}$ / $\sigma_{\rm gnss}$     &  $1$~s /  $5^{\circ}$  \\
        \bottomrule
    \end{tabular}
    \label{tab:sim-params}
    \vspace{-3mm}
\end{table}

To evaluate the performance of the proposed solution, we generate standard-compliant \ac{5g} \ac{pusch} \acp{dmrs} using the \textsc{5G Toolbox} of \textsc{matlab}~\cite{matlab5gtoolbox2023b}, and we use them as transmitted signals in a simulator that we developed in Python. We transmit $1$ \ac{dmrs} per slot. In the simulator, each reflector is represented by a point moving in space with a \ac{rcs} $\rho\sim\textrm{Lognormal}(\mu_{\rm rcs}, \sigma_{\rm rcs}^2)$ that follows a log-normal distribution, where $\mu_{\rm rcs}$ and $\sigma_{\rm rcs}^2$ are the mean and variance of $\log(\rho)$, to account for \ac{rcs} oscillations. 
The channel follows an \ac{awgn} model with a noise spectral density of $-174$~dBm/Hz (corresponding to a thermal noise temperature of $290$~K), and the \ac{tx} uses a power of $200$~mW, in line with \ac{3gpp} specifications~\cite [\S6.2]{3gpp38101-1}.
At the  \ac{rx} side, a least-squares channel estimation is performed using the received \ac{dmrs} symbols, and the subcarriers with no \ac{dmrs} are filled via linear interpolation. Then, \ac{cfo} and \ac{po} are added to the \ac{rx}.
From the \ac{cir} estimated at the \ac{rx}, the phases of the \ac{los}, target, and scatterer (when available) peaks are extracted and fed to our algorithm, along with the corresponding \acp{aod} at the \ac{tx}, whose error is modeled as an additive Gaussian noise $\mathcal{N}(0, \sigma_{\alpha}^2)$.
\tab{tab:sim-params} summarizes the simulation parameters.
Scatterers are available during on--off bursts of consecutive frames (\secref{sec:ref-scenario}).

\subsection{Simulation Results}
\label{sec:sim-results}

The proposed method is numerically evaluated in terms of normalized absolute estimation error, defined as $\varepsilon_{f_{\rm d}} = | f_{\rm d}[k] - \hat{f}_{\rm d}[k] | / | f_{\rm d}[k] |$, where $\hat{f}_{\rm d}[k]$ is the estimate of the true target Doppler ${f}_{\rm d}[k]$, and analogously for the \ac{tx} speed $\varepsilon_{v}$ and orientation $\varepsilon_{\eta}$. In the obtained estimates, the fusion ties the \ac{tx} kinematics to the target estimate, therefore, any residual $(v,\eta)$ error leaks through the self-motion compensation $g_{m}(v,\eta)$ of \eq{eq:g-function} and surfaces as an $f_{\rm d}$ bias.

\begin{figure}[t]
    \centering
    \subfloat[Varying \ac{aod} noise.\label{fig:fd-aod}]{\includegraphics[width=0.48\linewidth]{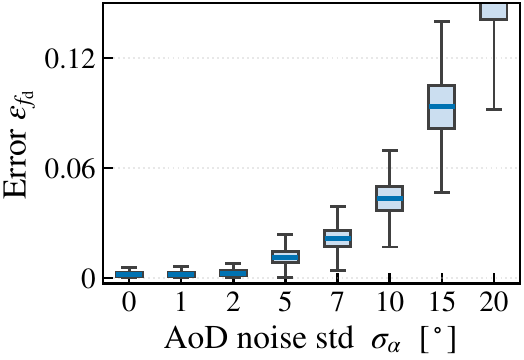}}
    \hfill
    \subfloat[Varying \ac{tx} and target speed.\label{fig:fd-speeds}]{\includegraphics[width=0.48\linewidth]{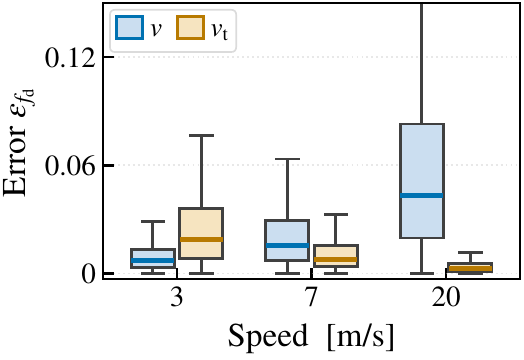}}
    \caption{Normalized target Doppler error $\varepsilon_{f_{\rm d}}$ vs. (a) \ac{aod} noise $\sigma_{\alpha}$ and (b) the \ac{tx} speed $v$ and the target speed $v_{\trg}$.}
    \label{fig:aod-speeds}
    \vspace{-0.3cm}
\end{figure}
\subsubsection{Impact of the AoD noise}

We first vary the \ac{aod} noise, which is critical since the angular information collapses the path geometry onto $\eta$, by varying  $\sigma_{\alpha}$ as in \fig{fig:fd-aod}. As expected, the error grows with $\sigma_{\alpha}$: small values ($\leq 2^{\circ}$) cause a modest degradation (median error below $0.5\%$ of the true Doppler), while larger ones impact the results significantly, with medians of $4.3\%$ and $15\%$ at $10^{\circ}$ and $20^{\circ}$, respectively.  
In the following, we fix $\sigma_{\alpha} = 5^{\circ}$ (median error $1.1\%$) as a default, compatible with the accuracy of the
\ac{5g} beam-management procedures.


\subsubsection{Impact of the TX and Target speeds}

We then explore the \ac{tx} and target speeds, considering $\{3, 7, 20\}$~m/s, representative of a pedestrian, a cyclist, and a car, respectively. The \ac{tx} speed has a marked effect, shown in \fig{fig:fd-speeds}. The median error grows from
$0.7\%$ to $5\%$, nearly proportional to $v$, as expected since the \ac{tx}-induced Doppler to be compensated scales as $v/\lambda$ by \eq{eq:tx-doppler}. The target speed, instead, leaves the error on $f_{\rm d}$ unchanged in absolute way, as $f_{\rm d}$ enters the target row linearly through the state-independent coefficient $2\pi T$, so its magnitude does not affect the estimation accuracy. Consequently, the normalized error reduces as the true Doppler grows (\fig{fig:fd-speeds}, medians of $1.9\%$, $0.8\%$, and $0.4\%$).

\begin{figure}[t]
    \centering
    \subfloat[Varying IMU grade.\label{fig:imu_all}]{\includegraphics[width=0.48\linewidth]{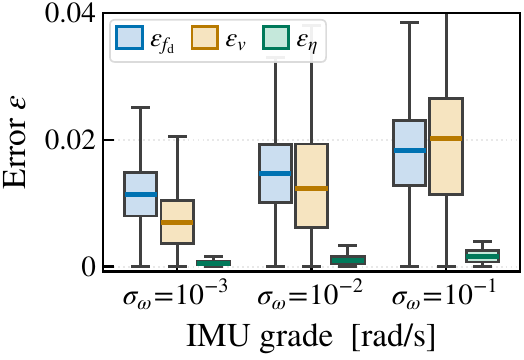}}    
    \hfill
    \subfloat[Varying scatterer visibility.\label{fig:fd-svis}]{\includegraphics[width=0.48\linewidth]{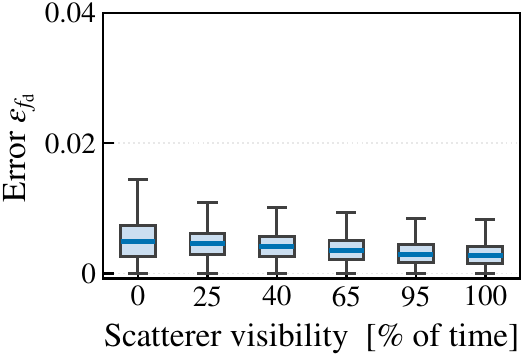}}
    \caption{(a) Normalized errors vs. the \ac{imu} grade: the target Doppler ($\varepsilon_{f_{\rm d}}$) together with the \ac{tx} speed ($\varepsilon_{v}$) and orientation ($\varepsilon_{\eta}$). (b) Normalized target Doppler error $\varepsilon_{f_{\rm d}}$ vs scatterer visibility with $\sigma_{\alpha} = 2^{\circ}$.}
    \label{fig:imu-all}
    \vspace{-0.3cm}
\end{figure}

\subsubsection{Impact of the IMU grade}

We test the three \ac{imu} grades $\sigma_{\omega} \in \{10^{-3}, 10^{-2}, 10^{-1}\}$~rad/s, with the remaining noise terms
scaled as in \secref{sec:imu-model}.
As expected, the error grows with the grade, but it does so slowly. Over two decades of $\sigma_{\omega}$ the medians move from $1.1\%$ to $1.8\%$ on $f_{\rm d}$, from $0.7\%$ to $2.0\%$ on $v$, and from $0.05\%$ to $0.17\%$ on $\eta$, as shown in \fig{fig:imu_all}. Even for the worst grade, the \ac{ekf} keeps the median heading error at $\approx0.5^{\circ}$, whereas the same unit integrated in open loop drifts by $\approx7^{\circ}$ over the $1$~s observation window: it is the fusion, and not the sensor grade, that sets the accuracy on the \ac{tx} motion, and a smartphone-class unit ($\sigma_{\omega} = 10^{-2}$~rad/s) is therefore sufficient for the proposed solution.

\subsubsection{Impact of the frame period}

\begin{figure}[t]
    \centering
    \subfloat[Varying frame period.\label{fig:fd-T}]{\includegraphics[width=0.48\linewidth]{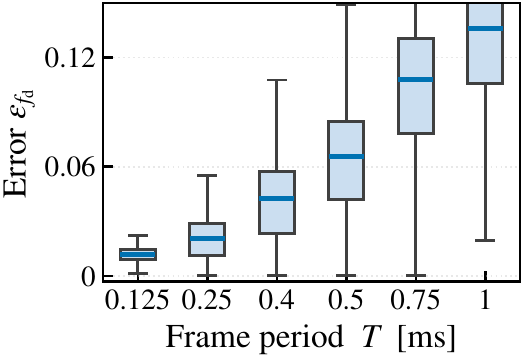}}
    \hfill
    \subfloat[ISAC and IMU baselines.\label{fig:isac_comp}]{\includegraphics[width=0.48\linewidth]{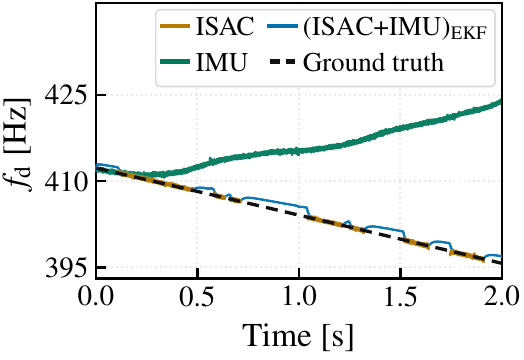}}
    \caption{(a) Normalized target Doppler error $\varepsilon_{f_{\rm d}}$ vs. frame period $T$ and (b) Target Doppler estimation of the proposed solution compared against the \ac{isac} and \ac{imu} baseline.}
    \label{fig:T-svis}
        \vspace{-0.3cm}
\end{figure}

We assess the frame period, varying $T\in \{0.125, 0.25, 0.4, 0.5, 0.75, 1\}$~ms (\fig{fig:fd-T}). The median $\varepsilon_{f_{\rm d}}$ grows monotonically with $T$, from $1.2\%$ at the base slot to $13.6\%$ at $1$~ms. The degradation is carried by the \ac{tx} kinematics, which reach $f_{\rm d}$ through $g_{m}(v,\eta)$. Although averaging over multiple phase measurements help, but two effects act in opposite direction. \emph{(i)} the number of phase updates within the window scales as $1/T$, so the state is corrected eight times less often at $1$~ms and the \ac{imu} errors are propagated in open loop over proportionally longer intervals, and \emph{(ii)} since \sys{eq:phase-row} requires the same scatterer at both ends of the $T$-long interval, the visibility bursts shorter than $T$ are never sampled,
which lowers the share of frames carrying a scatterer row and increases the mean interval between two re-anchoring instants.

\subsubsection{Impact of scatterer visibility}

We assess the performance by varying the fraction of time the scatterers are present during the simulation as $\{0, 25, 40, 65, 95, 100\}\%$ (\fig{fig:fd-svis}), adopting $\sigma_{\omega} = 10^{-1}$~rad/s for the \ac{imu}. We keep $N_{\rm s} = 2$, the minimum requirement to solve the system in \eqref{eq:phase-row} as in \cite{ventura2024bistatic, ventura2025asymov}. 

We group visible frames into bursts because the time domain phase differencing needs at least two consecutive visible frames to form a scatterer measurement; an isolated one contributes none. The filter produces estimates at every visibility level, including $0\%$. The error decreases monotonically with increased visibility, the median $\varepsilon_{f_{\rm d}}$ dropping from $0.7\%$ at $0\%$ (\ac{imu} alone) to $0.4\%$ at full visibility.
Note that, Visibility below $100\%$ yields no solution in the existing methods~\cite{ventura2024bistatic,
ventura2025asymov}.
 With a $\sigma_{\omega} = 10^{-2}$~rad/s unit, the \ac{imu} drifts so little between bursts that the scatterers would have essentially nothing left to correct. The solution thus grows more robust with scatterer availability, retaining useful accuracy even with intermittent or absent anchors.

\subsubsection{Comparison against the ISAC and IMU baseline}We finally compare our method against the \ac{isac} solution of~\cite{ventura2024bistatic, ventura2025asymov}, which recovers the same unknowns from the target and the two scatterer paths, and against the \ac{imu} solution. \ac{imu} solution here refers to the fusion of \ac{imu}-\ac{isac} with $0$ scatterers all the time and no \ac{ekf} integration, see \fig{fig:isac_comp}. We test the considered methods for $2$ s, for a scatterer visibility of $50$\% with unavailability periods ranging from $50$ to $500$~ms and a time varying target Doppler frequency. When scatterers are available, and thus \eqref{eq:phase-row} can be solved, the \ac{isac} baseline is accurate, but it provides no solution when the scatterers are absent. In contrast, the \ac{imu} track is always available, but it continuously drifts from the ground truth unless reset by an external source, i.e., the \ac{imu} initialization in \secref{sec:imu-model}. On the other hand, the proposed \ac{ekf}- based solution never stops (unlike \ac{isac}) yet provides accurate estimates (unlike \ac{imu}).

\section{Conclusions}
\label{sec:conclusions}

In this paper, we proposed a method to {\it jointly} estimate the Doppler frequency of a moving (passive) target, the transmitter speed and its orientation in clock asynchronous and bistatic mobile \ac{isac} systems. The approach uses extended Kalman filtering, by fusing the readings from a noisy inertial sensor at the transmitter with wireless channel estimates. Simulation results demonstrate that it is {\it robust}, seamlessly working in scenarios where the presence of scatterers is highly intermittent or absent over extended time periods; a condition where prior \ac{isac} solutions do not work. For typical settings, the median Doppler error remains below $1.2\%$ of the true target Doppler, considering a smartphone-grade \ac{imu} ($\sigma_{\omega} = 10^{-2}$~rad/s) and a 5G-NR compliant angle-of-departure noise of $5^{\circ}$. 
\bibliographystyle{IEEEtran}
\bibliography{IEEEabrv,refs}

\end{document}